\documentclass[tikz,11pt,english,a4paper]{article}
\usepackage{float}
\usepackage{jcappub}
\usepackage[utf8]{inputenc}

\usepackage{bm}
\usepackage{algorithm}
\usepackage{nicefrac}

\usepackage[noend]{algpseudocode}
\makeatletter
\def\BState{\State\hskip-\ALG@thistlm}
\makeatother

\usepackage{fontawesome5}
\usepackage{hyperref}
\makeatletter
\newcommand{\github}[1]{%
   \href{#1}{\faGithub}%
}
\makeatother

\usepackage{lmodern}
\newcommand{\dd}{\mathrm d}

\newcommand{\connect}{\textsc{connect}}
\newcommand{\class}{\textsc{class}}
\newcommand{\camb}{\textsc{camb}}
\newcommand{\montepython}{\textsc{MontePython}}
\newcommand{\tf}{TensorFlow}

\usepackage{siunitx}
\DeclareSIUnit \parsec {pc}

\usepackage[T1]{fontenc}

\usepackage{soul}

\usepackage[edges]{forest}
\definecolor{folderbg}{RGB}{124,166,198}
\definecolor{folderborder}{RGB}{110,144,169}
\newlength\Size
\tikzset{%
  folder/.pic={%
    \filldraw [draw=folderborder, top color=folderbg!50, bottom color=folderbg] (-1.05*\Size,0.2\Size+5pt) rectangle ++(.75*\Size,-0.2\Size-5pt);
    \filldraw [draw=folderborder, top color=folderbg!50, bottom color=folderbg] (-1.15*\Size,-\Size) rectangle (1.15*\Size,\Size);
  },
  file/.pic={%
    \filldraw [draw=folderborder, top color=folderbg!5, bottom color=folderbg!10] (-\Size,.4*\Size+5pt) coordinate (a) |- (\Size,-1.2*\Size) coordinate (b) -- ++(0,1.6*\Size) coordinate (c) -- ++(-5pt,5pt) coordinate (d) -- cycle (d) |- (c) ;
  },
}
\forestset{%
  declare autowrapped toks={pic me}{},
  declare boolean register={pic root},
  pic root=0,
  pic dir tree/.style={%
    for tree={%
      folder,
      font=\ttfamily,
      grow'=0,
    },
    before typesetting nodes={%
      for tree={%
        edge label+/.option={pic me},
      },
      if pic root={
        tikz+={
          \pic at ([xshift=\Size].west) {folder};
        },
        align={l}
      }{},
    },
  },
  pic me set/.code n args=2{%
    \forestset{%
      #1/.style={%
        inner xsep=2\Size,
        pic me={pic {#2}},
      }
    }
  },
  pic me set={directory}{folder},
  pic me set={file}{file},
}

\begin{document}


\title{Fast and effortless computation of profile likelihoods using CONNECT}

\author[a]{Andreas Nygaard,}
\author[a]{Emil Brinch Holm,}
\author[a]{Steen Hannestad,}
\author[a]{and Thomas Tram}

\affiliation[a]{Department of Physics and Astronomy, Aarhus University,
 DK-8000 Aarhus C, Denmark}

\emailAdd{andreas@phys.au.dk}
\emailAdd{ebholm@phys.au.dk}
\emailAdd{steen@phys.au.dk}
\emailAdd{thomas.tram@phys.au.dk}

\abstract{
	The frequentist method of profile likelihoods has recently received renewed attention in the field of cosmology. This is because the results of inferences based on the latter may differ from those of Bayesian inferences, either because of prior choices or because of non-Gaussianity in the likelihood function. Consequently, both methods are required for a fully nuanced analysis. However, in the last decades, cosmological parameter estimation has largely been dominated by Bayesian statistics due to the numerical complexity of constructing profile likelihoods, arising mainly from the need for a large number of gradient-free optimisations of the likelihood function. 
	
	In this paper, we show how to accommodate the computational requirements of profile likelihoods using the publicly available neural network framework \connect{} together with a novel modification of the gradient-based \textit{basin-hopping} optimisation algorithm. Apart from the reduced evaluation time of the likelihood due to the neural network, we also achieve an additional speed-up of 1--2 orders of magnitude compared to profile likelihoods computed with the gradient-free method of \textit{simulated annealing}, with excellent agreement between the two. This allows for the production of typical triangle plots normally associated with Bayesian marginalisation within cosmology (and previously unachievable using likelihood maximisation because of the prohibitive computational cost). We have tested the setup on three cosmological models: the $\Lambda$CDM model, an extension with varying neutrino mass, and finally a decaying cold dark matter model. Given the default precision settings in \connect{}, we achieve a high precision in $\chi^2$ with a difference to the results obtained by \class{} of $\Delta\chi^2\approx0.2$ (and, importantly, without any bias in inferred parameter values) -- easily good enough for profile likelihood analyses.
	}

\maketitle

\section{Introduction}\label{sec:introduction}
In the last few decades, parameter inference in cosmology has traditionally been done using Bayesian statistics. In Bayesian parameter inference, the goal is to characterise the multidimensional posterior probability distribution. This is often done using Markov-chain Monte Carlo sampling. Subsequently, estimates and uncertainties on single parameters are obtained by integrating the multidimensional posterior distribution over all other parameters, a process called marginalisation~\cite{10.1214/18-BA1112}. This is the main reason for the popularity of Bayesian parameter inference in cosmology: all credible intervals and two-dimensional posterior distributions are readily computable once a representative sample of the multidimensional posterior has been obtained.

Marginalisation requires a way to associate (prior) probability to volumes of parameter space, so the marginalised posterior distributions and the credible intervals~\cite{lee1989bayesian} may sometimes be completely dominated by the choice of prior. This effect is sometimes referred to as volume effects, because the effect is associated with the prior volume being integrated. This can be seen as an advantage because it makes it easy to build e.g.\ theoretical prejudice into the statistical analysis. For instance, one may wish to penalise a model containing a parameter that needs to be extremely fine-tuned to provide a good fit to the data. However, given that we often do not know the true underlying model, it could very well be that the true underlying model is observationally equivalent to the one proposed, but in the true model, the equivalent parameter is not fine-tuned. Thus, if we penalise the proposed model a priori, we might fail to discover the true model.

If we want to avoid this trap, we may instead employ frequentist statistics. Broadly speaking, this simply entails that we substitute marginalisation for maximisation; instead of integrating out the extra parameters, we maximise the likelihood over these parameters. The resulting object is the \textit{profile likelihood}, which has recently gained increased popularity~\cite{Cruz:2023cxy,Herold:2022iib,Holm:2022kkd,Reeves:2022aoi,Campeti:2022vom,Gomez-Valent:2022hkb,Herold:2021ksg,SPIDER:2021ncy,Planck:2013nga}. From this profile likelihood, we may then derive \textit{confidence intervals}~\cite{illowsky2017introductory}, akin to the credible intervals in Bayesian statistics. The advantage of the profile likelihood is that it may reveal interesting features of the model that would be missed in the Bayesian approach, but the disadvantage is the difficulty and computational cost associated with the large number of multidimensional optimisations. Early papers on cosmological parameter inference have examples of both marginalisation (see e.g.\  \cite{Tegmark:2000db}) and profiling (see e.g.\ \cite{Lineweaver:1996hi,Hannestad:2000hc,Hannestad:2003xv}; see also \cite{hamann2012} for an early discussion of profiling versus marginalisation). However, the introduction of MCMC techniques in marginalisation \cite{Christensen:2001gj,Lewis:2002ah} led to their increasing dominance in the field because of their speed and simplicity.

Many of the computational problems of profile likelihoods are solved by the recent advancements of machine learning within the field of cosmology. Many different cosmological emulators of observables, using e.g.\ neural networks~\cite{Nygaard:2022wri, SpurioMancini:2021ppk, Gunther:2022pto, LINNA, Bonici:2023xjk} or Gaussian processes~\cite{Gammal:2022eob, Gunther:2023xhh, ibanez}, have emerged in recent years, and these all benefit from much faster evaluation times than ordinary Einstein--Boltzmann solver codes.

In this paper, we show how \connect{}~\cite{Nygaard:2022wri} can facilitate fast and accurate computation of one- and two-dimensional profile likelihoods at a tiny fraction of the cost of a more traditional approach. This performance enhancement derives both from the speed-up of evaluating the neural network instead of \class{}~\cite{Blas:2011rf} or \camb{}~\cite{Lewis:1999bs} but also because the gradient of the likelihood can be computed fast and accurately through automatic differentiation techniques. The paper is organised as follows: In section~\ref{sec:introduction}, we give an introduction; in section~\ref{sec:profile}, we introduce profile likelihoods and discuss the difference between profile likelihoods and marginalised posteriors; in section~\ref{sec:optimisation}, we give a brief overview of the optimisation algorithms we use; and in section~\ref{sec:implementation}, we provide some more practical details regarding the implementation. In section~\ref{sec:results}, we show, for the first time, triangle plots for the profile likelihood compared to the posterior distribution for the $\Lambda$CDM-model as well as for an extension with varying neutrino mass and degeneracy. We also show a profile likelihood of the decay constant in a decaying cold dark matter model as an example where the training data of the neural network itself suffers from large volume effects. Finally, we give our conclusions and future outlook in section~\ref{sec:conclusion}.

\section{Profile likelihoods and Bayesian inference}\label{sec:profile}
Given an  $N$-dimensional parameter space $\Theta$ where we are mainly interested in constraining the parameters of an $M$-dimensional subset $\Omega$, the profile likelihood $\mathcal{L}(\vec{\theta})$ of a likelihood function $\mathcal{L}(\vec{\theta}, \widetilde{\theta})$ on the full parameter space is obtained by maximising the dimensions not contained in the reduced space $\Omega$~\cite{pawitan},
\begin{align}\label{eq:pl_definition}
	\mathcal{L} (\vec{\theta}) = \max_{\widetilde{\theta}} \!\left(\mathcal{L}(\vec{\theta}, \widetilde{\theta})\right), \quad \vec{\theta} \in \Omega, \widetilde{\theta} \in \Theta \setminus \Omega,
\end{align}
where $\vec{\theta}$ represents a vector in the parameter subspace of interest and $\widetilde{\theta}$ a parameter vector in the subspace of $\Theta$ we maximise over. Parameters in the latter subspace are commonly referred to as nuisance parameters, and usually we are interested in either one-dimensional profile likelihoods, $M=1$, for one-parameter constraints, or two-dimensional profile likelihoods, $M=2$, to constrain the correlation between pairs of parameters. Since the profile likelihood~\eqref{eq:pl_definition} is obtained by maximising a subset of the parameter space, frequentist inference based on it is an optimisation problem (as opposed to Bayesian inference, which is a sampling problem). The main contributions of this paper are novel strategies for carrying out this optimisation.

From the profile likelihood, confidence intervals (or \textit{regions}, if $M>1$) can be obtained using the Neyman construction~\cite{Neyman:1937uhy}, where 68.27\% (95.45\%) confidence regions for $\vec{\theta}$ are defined implicitly by the regions $\Delta\chi^2 (\vec{\theta}) < 1.0 \ (4.0)$ with the definition
\begin{align}
	\Delta\chi^2 (\vec{\theta}) \equiv -2 \log \left( \frac{\mathcal{L}(\vec{\theta})}{\max\limits_{\vec{\theta}} \!\left(\mathcal{L} (\vec{\theta})\right)} \right).
\end{align}
The coverage of the intervals constructed using the Neyman method is exact whenever the profile likelihood is Gaussian or whenever there exists a reparametrisation in which it is. Alternative interval constructions, such as the Feldman-Cousins prescription~\cite{Feldman:1997qc}, are known to produce more accurate interval coverages, but since the focus of this paper is on the optimisation and not the interval construction, we employ the Neyman construction throughout.

\begin{figure}
	\includegraphics[width=\textwidth]{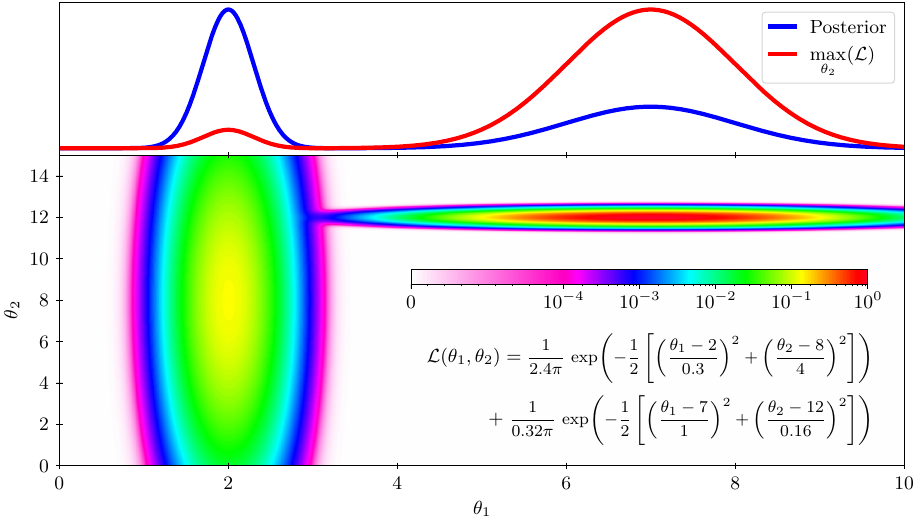}
	\caption{\textsl{The bottom panel shows the function value of the likelihood function written in the panel, while the top panel shows the resulting marginalised posterior and the profile likelihood, both scaled to a maximum of unity. The posterior is dominated by the shorter large Gaussian, while the profile is dominated by the taller small Gaussian. The two one-dimensional statistics thus reveal complementary information about the actual likelihood.}}
	\label{fig:postvsprof}
\end{figure}

As seen in the definition~\eqref{eq:pl_definition}, the profile likelihood is a maximum likelihood estimate in the reduced subspace and therefore inherits the invariance under reparametrisations of this subspace from the latter~\cite{pawitan}. This is in contrast to posterior distributions from the Bayesian inferences, which may change with the specific parametrisation. To illustrate the difference between the marginalised posterior distribution and the profile likelihood, we investigate a toy likelihood comprised of two Gaussian distributions with different normalisations and covariance matrices. The two Gaussians have only a slight overlap, as shown in figure~\ref{fig:postvsprof}, and the one with the largest maximum is much more narrow in the $\theta_2$ parameter. This makes the contribution of the taller peak to the posterior in the $\theta_1$ parameter much less when marginalising over the $\theta_2$ parameter in the case of uniform priors, even though the likelihood is actually larger in this peak. This shows how the greater volume of a less significant likelihood peak can dominate the marginalised posterior. When computing a profile likelihood in the $\theta_1$ parameter, the likelihood is optimised for each fixed value of $\theta_1$, and so the profile looks like a projection of the two-dimensional surface onto the one-dimensional $\theta_1$-axis. This, of course, shows the two peaks with their actual height differences.

The two different ways of representing the likelihood function come from the two different ways of interpreting probability in frequentist and Bayesian statistics. Neither method is more correct; they simply answer different questions. Frequentist inference answers the question of how we can choose a range in the parameter space based on the data such that the true value of the parameter will fall within the range in $(1-\alpha)\times100\%$ of the asymptotically often repeated realisations of the experiment. This range is called the confidence interval with the level of significance $\alpha$. Bayesian inference treats the true value of the parameter as a random parameter chosen from a distribution, and the question is which range of the parameter space we can choose so that we are $(1-\alpha)\times100\%$ certain that the true value has taken a value that lies within the range. This is called the credible interval with the level of significance $\alpha$. The posterior can therefore be thought of as a probability distribution, while the profile likelihood cannot. Ultimately, it is this difference in interpretation that manifests in the different constraints obtained from the two statistical paradigms.

\section{Optimisation of the likelihood function}\label{sec:optimisation}
An accurate and robust optimisation routine is crucial for profiling likelihoods, and this task can be difficult in certain situations. Optimisation routines typically require many function evaluations in order to perform well, and this is especially true if there is no prior knowledge of the cost function. In the case of using an Einstein--Boltzmann solver code like \class{}, the function takes on the order of 10 core seconds to evaluate (including both the Einstein--Boltzmann solver code and the likelihood codes), and this quickly adds up to large computation times. Certain optimisation methods make use of gradients, but Einstein--Boltzmann solver codes tend to be numerically unstable with respect to differentiation due to different approximation schemes toggling in different regions of parameter space. Furthermore, in order to construct profile likelihoods, it is essential that the global optimum be found. Therefore, global optimisation routines are essential. 

\subsection{Global optimisation}

The problem of global optimisation is numerically difficult. Methods like \textit{gradient descent}~\cite{5a25e804-97fa-3be4-ac27-50be05d4f28d} and the \textit{simplex algorithm}~\cite{pjm/1103044531} can get stuck in local optima, so this can be a problem whether or not one has access to gradients of the likelihood function. There exist, however, specific methods that can effectively search the parameter space in clever ways inspired by MCMC methods. One such method that has proven to be fairly efficient is \textit{simulated annealing}~\cite{doi:10.1126/science.220.4598.671}, which does not require gradients and is therefore a suitable choice when dealing with Einstein--Boltzmann solver codes. In short, this method searches the parameter space as any other MCMC but gradually lowers the sampling temperature while doing so. This makes the features of the likelihood landscape more profound over time, allowing the MCMC chain to eventually settle on the global optimum. The optimisation is highly dependent on the chosen temperature schedule and somewhat inefficient when the acceptance rate of the MCMC is small. Despite the individual simulated annealing optimisation being sequential, each point in a profile likelihood can be optimised in parallel, so one can therefore do profile likelihoods based on simulated annealing in reasonable time, although the number of evaluations needed for an inference of all parameters in a typical cosmological model, as well as their correlations, makes it unfeasible to do with an Einstein--Boltzmann solver code. 

Instead, we use a neural network from the \connect{} framework to speed up the evaluation of the observables. This means that the likelihood evaluation time is dominated by the comparison of theoretical observables to data. This is, however, a significant decrease in computational cost, and using the same method as before (simulated annealing), we can obtain profile likelihoods much quicker. 

We can even improve on the method now that we are using a neural network that does not suffer the same numerical instabilities with respect to differentiation as Einstein--Boltzmann solver codes do. The auto-differentiation of the \tf{}~\cite{tensorflow} framework lets us easily differentiate the neural network with respect to the input parameters, so we can now make use of gradient-based optimisation techniques. However, before that, we need to be able to auto-differentiate the likelihood calculation as well. We are seeking the derivative of the likelihood value $\mathcal{L}$ with respect to the cosmological input parameters $\vec{\theta}$, and the network only provides us with the derivative of the observables $\vec{\mathcal{O}}$ ($C_\ell$ spectra, power spectra, etc.) with respect to the cosmological parameters. By the chain rule, we therefore need the derivative of the likelihood calculation with respect to its input (the cosmological observables computed by the network),
\begin{align}\label{eq:diff1}
	\frac{\dd \mathcal{L}}{\dd \vec{\theta}} = \frac{\dd \mathcal{L}}{\dd \vec{\mathcal{O}}}\frac{\dd \vec{\mathcal{O}}}{\dd \vec{\theta}}\,.
\end{align}
Obtaining the derivative of the likelihood with respect to cosmological observables proves to be quite elaborate, but this is discussed further in section~\ref{sec:implementation}. 

Equipped with auto-differentiation all the way from cosmological input parameters to the likelihood value, one can now make use of gradient-based methods. Relying solely on these can be a problem since gradient-based methods are often unable to explore other optima than the closest. A global optimisation method that can circumvent this issue is the \textit{basin-hopping}~\cite{doi:10.1021/jp970984n} algorithm. This method is reminiscent of simulated annealing, but with the addition of a local optimisation in each step. The method therefore "hops" between different local optima, or "basins", instead of hopping between random points. This reaches convergence much quicker since fewer steps are needed due to the local optimiser always placing each point at an optimum. This method is very dependent on a good local optimisation method; otherwise, it will reduce to simulated annealing in the limit of a trivial optimiser. One drawback of this optimisation method is that one is not guaranteed to find the global optimum given the stochasticity of the optimiser~\cite{Zhou2019}, and indeed there is a probability of convergence at a suboptimal point (see appendix~\ref{app:addfix}). In our case, however, this is a quite small probability due to the smoothness of the neural networks, and it can be heavily decreased through precision settings.

\subsection{Local optimisation}
A local optimisation method should, in our case, take only a few steps in order to reach the optimum. Gradient-based methods are obviously suitable for this, but the simplest of such methods, gradient descent, does not perform well in this regard. The reason for this is that the step size will decrease with the slope, and a lot of steps will be taken close to the optimum due to the vanishing gradient. Some parts of the likelihood function can even be close to flat, and this requires a lot of steps by the gradient descent optimiser. A better choice could be to also use second-order derivatives, which would mean that we could almost guess the correct optimum after one evaluation if the likelihood landscape is close to Gaussian. The second-order derivatives are, however, not easy to get in our case since the \tf{} graph of this computation becomes too large to handle. An appropriate compromise could be a pseudo-second-order method like the \textit{Levenberg--Marquardt}~\cite{83b09f23-b20e-3617-8f72-24765b713f7b,doi:10.1137/0111030} or \textit{Broyden--Fletcher--Goldfarb--Shanno}~\cite{Broyden1970,10.1093/comjnl/13.3.317,35d0019d-775a-3628-b0b4-67be112e346b,e3177091-3094-3792-9d61-0ab445735ddb} (BFGS) algorithms. These use first-order derivatives to approximate the Hessian in order to quickly locate the nearest optimum. \tf{} has an implementation of BFGS, so this has been chosen as the local optimiser to use with the global basin-hopping method.

\section{Implementation}\label{sec:implementation}
\subsection{Auto-differentiability}
In order for us to use gradient-based methods, we need auto-differentiation at each step in the evaluation of the likelihood function. We use the built-in \textit{reverse mode differentiation} of \tf{}, and this requires every operation during the evaluation to be written with \tf{} syntax in order for them to be auto-differentiable. Many popular likelihood codes, such as the full Planck 2018 likelihood~\cite{Planck:2019nip} (including low-$\ell$ $TT$, low-$\ell$ $EE$, high-$\ell$ $TT+TE+EE$, and lensing), are complex and tedious to translate to \tf{} syntax, so as of now only the Planck lite likelihood has been translated, first to \textsc{Python} by Ref.~\cite{Prince:2019hse} and then to \tf{} by Ref.~\cite{SpurioMancini:2021ppk}. We have altered the \tf{} version from Ref.~\cite{SpurioMancini:2021ppk} to accommodate neural networks from \connect{} and this involved another rather tedious task of interpolation. \connect{} only computes the same $\ell$-grid that \class{} does, and this is more than an order of magnitude fewer points than required by the likelihood code. We have therefore implemented a cubic spline interpolation routine in the likelihood code since no suitable interpolation method is implemented in \tf{}. This has to be written not only in \tf{} syntax but also using only differentiable operations (some functionality in \tf{} is not auto-differentiable). This adds an extra layer of computation to equation~\eqref{eq:diff1},
\begin{align}\label{eq:diff2}
	\frac{\dd \mathcal{L}}{\dd \vec{\theta}} =   \underbrace{\frac{\dd \mathcal{L}}{\dd \vec{C}_{\ell}^{[2508]}}}_\text{Likelihood code}      \overbrace{\frac{\dd \vec{C}_{\ell}^{[2508]}}{\dd \vec{C}_{\ell}^{[100]}}}^\text{Interpolation}      \underbrace{\frac{\dd \vec{C}_{\ell}^{[100]}}{\dd \vec{\theta}}}_\text{Neural network} \,,
\end{align}
where $\vec{C}_\ell$ is a vector of $C_\ell$ values, which is the observable used by the likelihood code in our case, and the number in square brackets tells the number of $C_\ell$ values. One could simply emulate all 2508 values needed by the code, but this is a computational waste when training the network since all the information is contained in only the 100 values that \class{} actually calculates~\cite{Nygaard:2022wri} (the number of points calculated by \class{} differs based on input and precision settings). Since each of the computational steps in equation~\eqref{eq:diff2} is now differentiable, the total derivative of the likelihood with respect to the cosmological parameters can be used for optimisation purposes. A single function evaluation of the gradients takes on the order of $\sim$$10^{-2}$ seconds and is approximately twice as expensive as evaluating just the likelihood itself (using the neural network, the interpolation, and the data comparison).

\subsection{Ensemble basin-hopping}

When training a neural network with \connect{}, the training data is gathered using multiple MCMC runs in an iterative manner~\cite{Nygaard:2022wri}, where each iteration uses an MCMC sampler to gather data using the neural network from the previous iteration. This means that we already have a covariance matrix and an estimate for the best-fit point as a starting point. This is a great help when doing the actual maximum likelihood optimisation, since the likelihood landscape is roughly known beforehand. We can use this information to slightly modify the standard basin-hopping algorithm to accommodate this additional information. Steps are normally taken sequentially, but since we have a covariance matrix and a best fit estimate, we can construct a proposal distribution and draw multiple points from it at once. These can then all be locally optimised in parallel, thus exploring different local optima simultaneously. We can then centre a new proposal distribution around the best point of the ensemble of optimised points and lower the sampling temperature. Repeating this allows us to home in on the global optimum much faster due to the parallelisability. The altered algorithm is sketched below\footnote{A similar approach can be found in Ref.~\cite{inproceedings}, where multiple individual workers perform independent basin-hopping optimisations but exchange information about optimal starting points through a master process.}:\\

\begin{algorithmic}
\State $Cov(\vec{\theta}) =$ covariance matrix from MCMC
\State $\vec{b} \hspace{2.65em}=$ estimate of best-fit point
\State $opt(\vec{\theta})  \hspace{0.45em}=$ local optimiser                                                                                                 \hspace{15em}(\textit{returns value and position})
\State $p(\vec{\theta})  \hspace{1.3em}=$ Gaussian distribution from $Cov(\vec{\theta})$ and $\vec{b}$                            \hspace{3.5em}(\textit{proposal distribution})
\State $T  \hspace{2.35em}= 1.0$                                                                                                                                          \hspace{20.2em}(\textit{sampling temperature})
\State $T_{\rm min}  \hspace{1.15em}=$ minimal temperature                                                                                              \hspace{12.15em}(\textit{works as a tolerance})
\State $N  \hspace{2.16em}=$ number of points in ensemble

\While{$T>T_{\rm min}$}
\State $\mathcal{L}_{\rm array} = {\rm zeros}(N)$
\State $P_{\rm array} \hspace{0.05em}= {\rm zeros}(N)$

\For{$i\;\textbf{in}\;\textrm{range}(N)$}                                                                                                                                    \hspace{15.1em}(\textit{parallelisable})
\State $\vec{X} \hspace{6.48em}=$ point drawn from $p(\vec{\theta})$
\State $\mathcal{L}_{\rm array}[i],\, P_{\rm array}[i] = opt(\vec{X})$ 
\EndFor

\State $\mathcal{L}_{\rm min} = {\rm min}(\mathcal{L}_{\rm array})$                                                                                      \hspace{15.9em}(\textit{new best value})
\State $\vec{b} \hspace{1.61em}=$ element in $P_{\rm array}$ corresponding to $\mathcal{L}_{\rm min}$                          \hspace{3.43em}(\textit{new best-fit point})
\State $T \hspace{1.31em}= T/2$                                                                                                                                            \hspace{19.13em}(\textit{reduce the temperature})
\State $p(\vec{\theta})\hspace{0.26em}=$ proposal distribution from $Cov(\vec{\theta})\cdot T$ and $\vec{b}$

\If{the majority of the ensemble finds the same optimum}
\State break from while loop

\EndIf

\EndWhile
\State $\mathcal{L}_{\rm min}$ is now the optimised function value and $\vec{b}$ is the best-fit point

\end{algorithmic}

\subsection{Constraints on parameter space}
Profile likelihoods are not subject to any priors as marginalised posteriors are, but there might still be benefits to confining the optimisation within certain ranges in the parameter space due to physical constraints. An example here could be the mass of a particle, which must be positive. When computing profile likelihoods using Einstein--Boltzmann solver codes and simulated annealing, this will never be an issue since the code will raise a computation error in such a case, and the likelihood code will then return a very low likelihood such that no point beyond this physical boundary will be accepted. Neural networks will, however, always produce output based on any input of the correct type and dimensionality, so they might learn some behaviour linked to a certain parameter as it decreases. It is then able to extrapolate, and the result might be a good fit to the data either by chance or because of this extrapolation. This is exactly the case for the model with a varying neutrino mass used as a test case in this paper. It is therefore beneficial to confine our optimisation within given parameter ranges. A variant of the BFGS algorithm dubbed BFGS-B~\cite{doi:10.1137/0916069} performs the optimisation within a confined box in the parameter space. This is very useful in our case, but unfortunately it is not implemented in \tf{}. A solution is therefore to introduce a smooth and differentiable penalty function that penalises the likelihood when evaluating a point outside of the box. We have chosen a very steep exponential function that increases depending on the distance from the boundary, and this ensures that gradients are still meaningful in a way such that any evaluation outside the box will send the optimiser back inside the box. A proper implementation of BFGS-B might be beneficial in the future.

\subsection{Workflow}
When doing profile likelihoods with \connect{}, there are a few steps. First of all, one needs to gather training data and train a neural network with a specific cosmological model implemented in \class{}. Then, it is a good idea to run a normal MCMC with the neural network in order to have a good covariance matrix along with Bayesian inference for comparison. The covariance matrix from the gathered training data can be used instead\footnote{The choice of covariance matrix does not significantly impact the result as long as the initial proposal distribution is wide enough to encapsulate the best-fit region.}. Then one needs to choose at which points in the parameter space to optimise for both one- and two-dimensional profile likelihoods; the idea is to sample with more resolution where the likelihood function varies significantly as well as near its maximum. This can be tricky if one does not know any features of the likelihood function of the particular cosmological model, and this is another reason for doing an MCMC run with the neural network beforehand. When the posteriors are known, a good initial guess is that the profile likelihood will somewhat resemble them. This is exactly true when there are no volume effects and only flat priors are used, but in any case, it is reasonable to assume that at least some features of the profile likelihood will overlap with the features of the posterior. In the case of one-dimensional profile likelihoods, one can often get away with simply choosing a set of equally spaced points. In the case of two-dimensional profile likelihoods, it is not as straightforward, but a good guess is to use the points of the histograms from which the posteriors are computed. Any non-zero bin in these histograms corresponds to a region where the MCMC has accepted points, and so we can choose these bin centres as the points in our two-dimensional profiles (see appendix~\ref{app:2d}). If the computed profiles seem to not be best represented by these points, a routine to manually add points by clicking in the plots has been included (see appendix~\ref{app:addfix}). A full automatisation of which points to choose is beyond the scope of this work but should be further investigated. 

\section{Results and discussion}\label{sec:results}
In order to test the performance of the optimisation routine, we have chosen three cosmological models as examples: $\Lambda$CDM, massive neutrinos, and decaying cold dark matter. These three models each have features that are useful for highlighting different problems and their corresponding solutions.

\subsection{$\Lambda$CDM}
The posterior distribution of the standard $\Lambda$CDM model is almost perfectly Gaussian under standard CMB data~\cite{Planck:2019nip} and therefore has no volume effects. In this case, we expect to see the profile likelihoods coinciding perfectly with the posteriors in both the one- and two-dimensional plots. We did not train a specific $\Lambda$CDM neural network with \connect{} since this will be contained in the neural networks of the other two models. We have therefore used the same network as for the massive neutrinos model, where the parameters $m_{\rm ncdm}$ and ${\rm deg}_{\rm ncdm}$ were fixed to $0.06$ and $1.0$, respectively, in order to match a value of the effective number of degrees of freedom of $N_{\rm eff}=3.046$. When fixing the two model-specific parameters, the rest of the network behaves like a $\Lambda$CDM network trained with these two parameters fixed at the same values.

\begin{figure}[t]
	\centering
	\includegraphics[width=\textwidth]{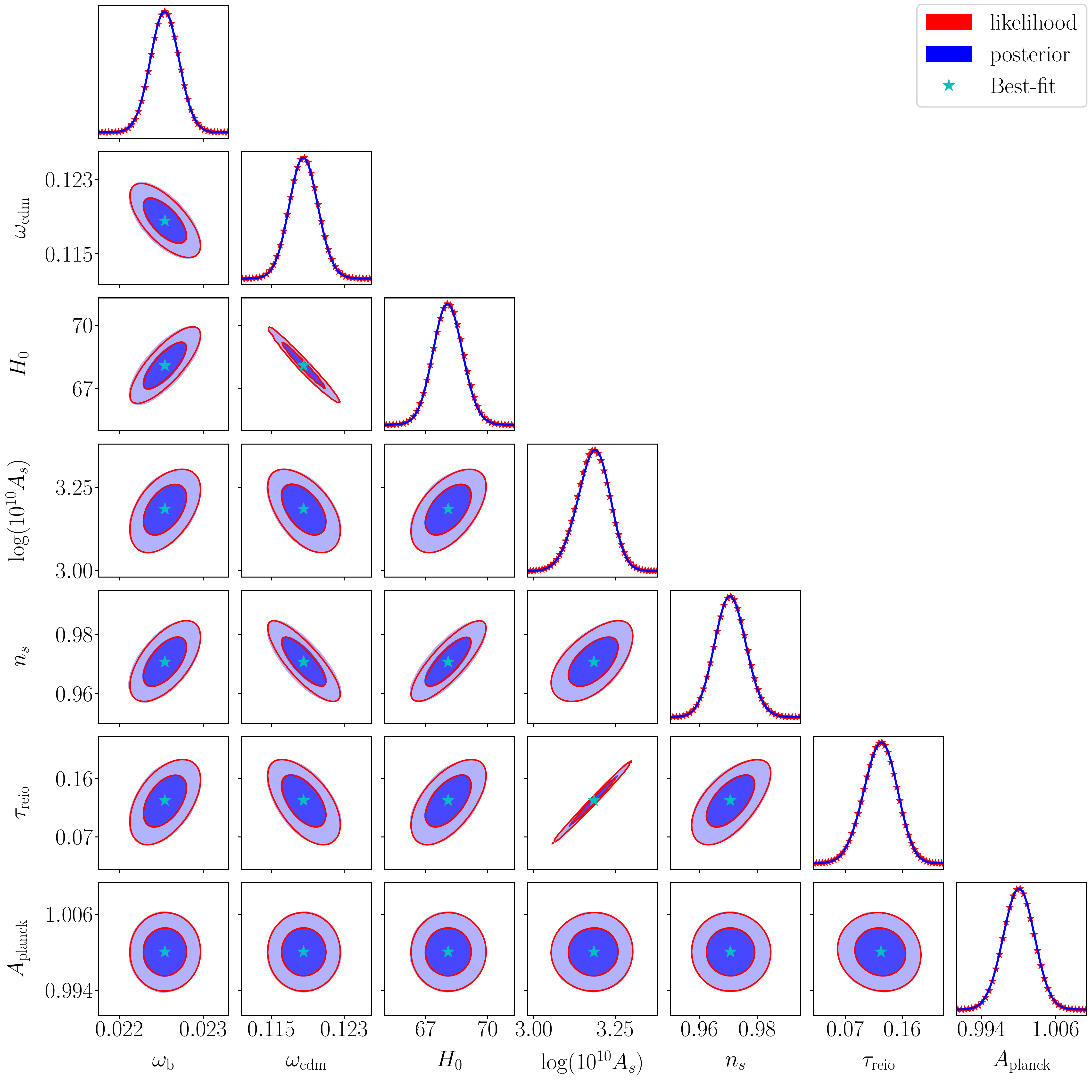}
	\caption{\textsl{Posteriors and profile likelihoods of the $\Lambda$CDM model. The blue filled contours and the blue lines on the diagonal are the posteriors from an MCMC run with the neural network for one and two dimensions, respectively, and the red contour lines and the red stars on the diagonal are the profile likelihoods for one or two dimensions, respectively. The cyan star marks the best-fit point from a global optimisation of the entire neural network.}}
	\label{fig:lcdm_tri}
\end{figure}

\begin{figure}[t]
	\centering
	\includegraphics[width=\textwidth]{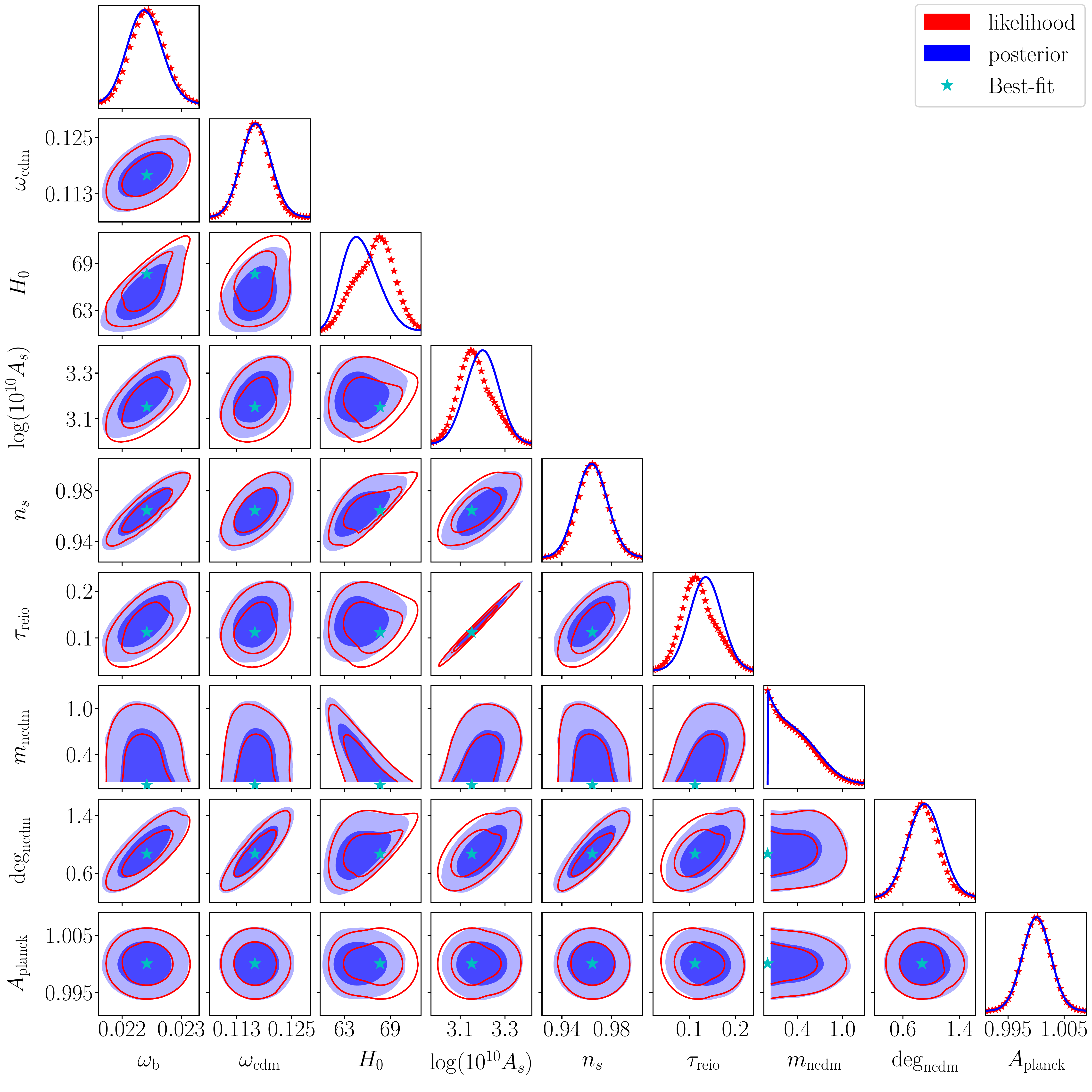}
	\caption{\textsl{Posteriors and profile likelihoods of the massive neutrinos model. The blue filled contours and the blue lines on the diagonal are the posteriors from an MCMC run with the neural network for one and two dimensions, respectively, and the red contour lines and the red stars on the diagonal are the profile likelihoods for one or two dimensions, respectively. The cyan star marks the best-fit point from a global optimisation of the entire neural network.}}
	\label{fig:ncdm_tri}
\end{figure}

Figure~\ref{fig:lcdm_tri} shows a full triangle plot of both posteriors and profile likelihoods for the $\Lambda$CDM model. The blue lines and filled contours are the one- and two-dimensional posteriors, respectively; the red stars and contour lines are the one- and two-dimensional profile likelihoods, respectively; and the cyan stars mark the best-fit point of the entire parameter space. The stars in the one-dimensional plots are the actual calculated points in the profile likelihoods, but for the two-dimensional plots, we only show the contour lines calculated from the computed points in the same way that the posterior contours are calculated from the histograms. The agreement between the posteriors and the profile likelihood is excellent, and since we would expect this to be true for a near-Gaussian likelihood, this validates our optimisation routine.

\subsection{Massive neutrinos}

Now that we have tested our optimisation on a simple likelihood without any volume effects, like in the $\Lambda$CDM model, we can move on to a model that actually contains volume effects. The inclusion of a variable neutrino mass, $m_{\rm ncdm}$, (two species are assumed to be massless) and the degeneracy, ${\rm deg}_{\rm ncdm}$, introduces new volume in the parameter space, and the likelihood is not Gaussian for the neutrino mass. We should therefore expect to see some differences between the posteriors and the profile likelihoods. Figure~\ref{fig:ncdm_tri} shows the posteriors and profile likelihoods of this model. The posteriors are again shown in blue, while the profile likelihoods are shown in red. We can clearly see differences between the posteriors and profile likelihoods now, with the profile likelihoods being shifted slightly compared to the posteriors. The most profound shifts are to higher values of $H_0$ and lower values of both $A_s$ and $\tau_{\rm reio}$, and it is interesting that the actual best-fit point (cyan stars) lies on the edge of the 68.27\% credible regions of the posteriors involving $H_0$, and the 95.45\% confidence regions of the profile likelihoods in $H_0$ seem to alleviate the Hubble tension significantly. Since we are only using the Planck lite likelihood, we cannot draw any reasonable conclusions based on this, but it really emphasises how the volume of the parameter space can impact parameter inference and that both a Bayesian analysis and a frequentist analysis should be performed in order to get the full picture. 

\begin{figure}[t]
	\centering
	\includegraphics[width=\textwidth]{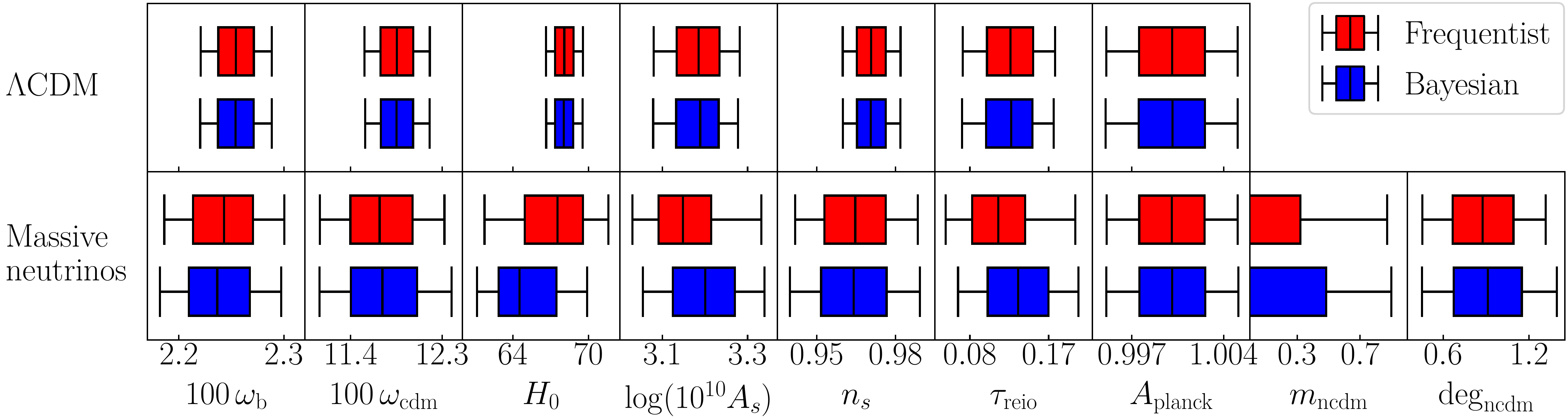}
	\caption{\textsl{68.27\% and 95.45\% confidence (credible) intervals found using frequentist (Bayesian) statistics for the two models, $\Lambda$CDM and massive neutrinos. We clearly see that the constraints from posteriors (blue) and profile likelihoods (red) are identical for $\Lambda$CDM (top panel) but differ somewhat for massive neutrinos (bottom panel). Along with the constraints, the best-fit point has been included as a centreline in the boxes.}}
	\label{fig:interval}
\end{figure}

The differences between a model with volume effects and one without are more apparent from figure~\ref{fig:interval}, where the 68.27\% and 95.45\% constraints from both the profile likelihoods and the posteriors are shown for each of the two cosmological models, the $\Lambda$CDM model and the massive neutrinos model. The additional volume of the massive neutrinos model not only broadens the constraints but also shifts some of the parameters, as previously discussed.

\begin{figure}[t]
	\centering
	\includegraphics[width=\textwidth]{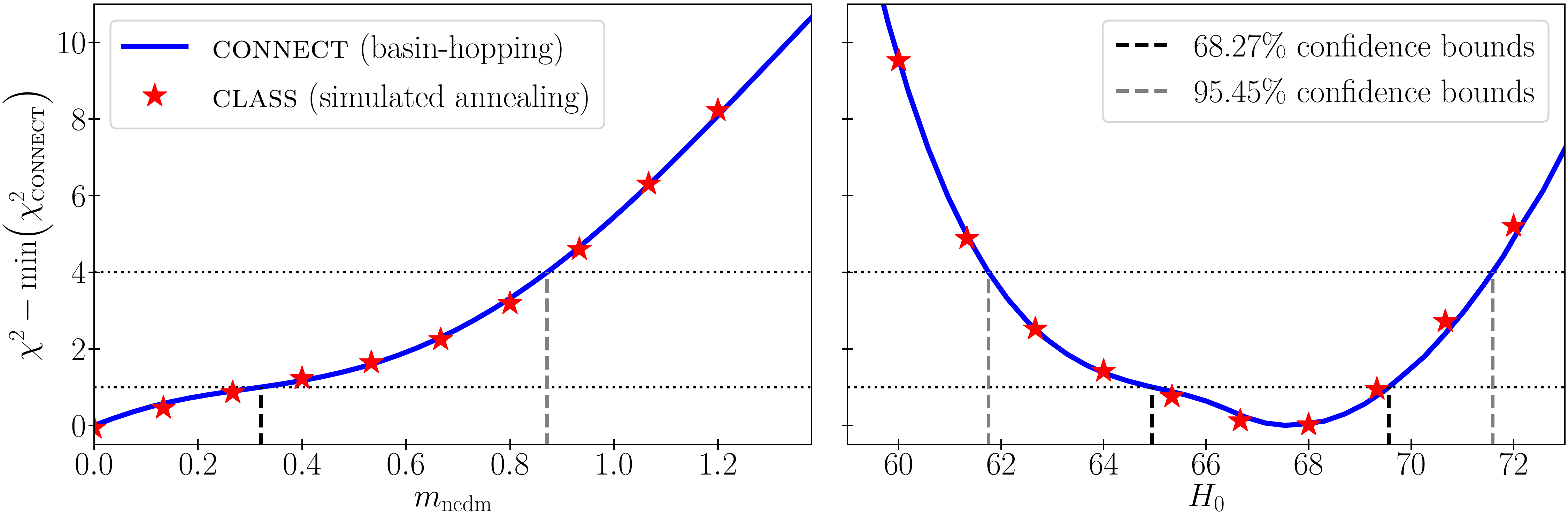}
	\caption{\textsl{Comparison between one-dimensional profile likelihoods using \class{} and \connect{} with simulated annealing and basin-hopping as optimisation routines, respectively. The profile likelihoods are in the parameters $m_{\rm ncdm}$ and $H_0$. The 68.27\% and 95.45\% confidence intervals are also shown.}}
	\label{fig:1d}
\end{figure}

In order to test the performance of the emulator, we compare the results with profile likelihoods obtained using \class{} and simulated annealing. These are shown in figure~\ref{fig:1d}, along with the 68.27\% and 95.45\% confidence intervals calculated from the \connect{} profile likelihood. We see a great agreement between \class{} and \connect{} and this further strengthens our confidence in the results obtained using \connect{} in figures~\ref{fig:lcdm_tri}~and~\ref{fig:ncdm_tri}.  

\subsection{Decaying cold dark matter}

In order to really put the framework and optimisation to the test, we have included the model of decaying cold dark matter (DCDM) with dark radiation (DR) as the decay product. This likelihood is notoriously difficult to sample and is also quite challenging when doing profile likelihoods~\cite{Holm:2022kkd}. With Planck lite data, the likelihood features a very slight peak for high values of the decay rate, $\Gamma_{\rm dcdm} $, with a height corresponding to only $\Delta\chi^2\approx0.5$. For this, we require much higher precision than for the other cosmological models, but achieving this turns out to be very difficult. The optimisation can only ever be as good as the neural network used by the optimisation routine, and a good network suitable for this particular optimisation requires a lot of training data around this subtle likelihood peak. As of now, we are limited by the fact that our training data for the neural network is generated by iteratively sampling from the posterior, and this makes it very hard to properly sample around the peak due to the large volume effects that the posterior is influenced by. It can, of course, be solved by sampling maybe an order of magnitude more points to use as training data (as well as training for more epochs and with a larger network architecture), but this becomes unfeasible in terms of training the network. 

Another approach would be to slice the parameter space at specific values for any parameters exhibiting such problems and train different networks for each value of this parameter. This also requires more computational resources, and if the goal is to have just the one-dimensional profile likelihood in the parameter in question ($\Gamma_{\rm dcdm}$ in our case), this is quite wasteful, and one might as well compute the profile likelihood with \class{} and simulated annealing. We can, however, use these individual networks for more, and so if we wanted a full triangle plot of profile likelihoods, we would just use the appropriate network for any point in a two-dimensional profile likelihood containing the sliced parameter and use all networks at once for any point in all other two-dimensional profile likelihoods and simply choose the lowest value of $\chi^2$. This adds a layer of complexity compared to using just a single network containing the whole parameter space, but if any two-dimensional profile likelihoods are needed, then this approach will be orders of magnitude less computationally expensive than using \class{} and simulated annealing.  

\begin{figure}[t]
	\centering
	\includegraphics[width=\textwidth]{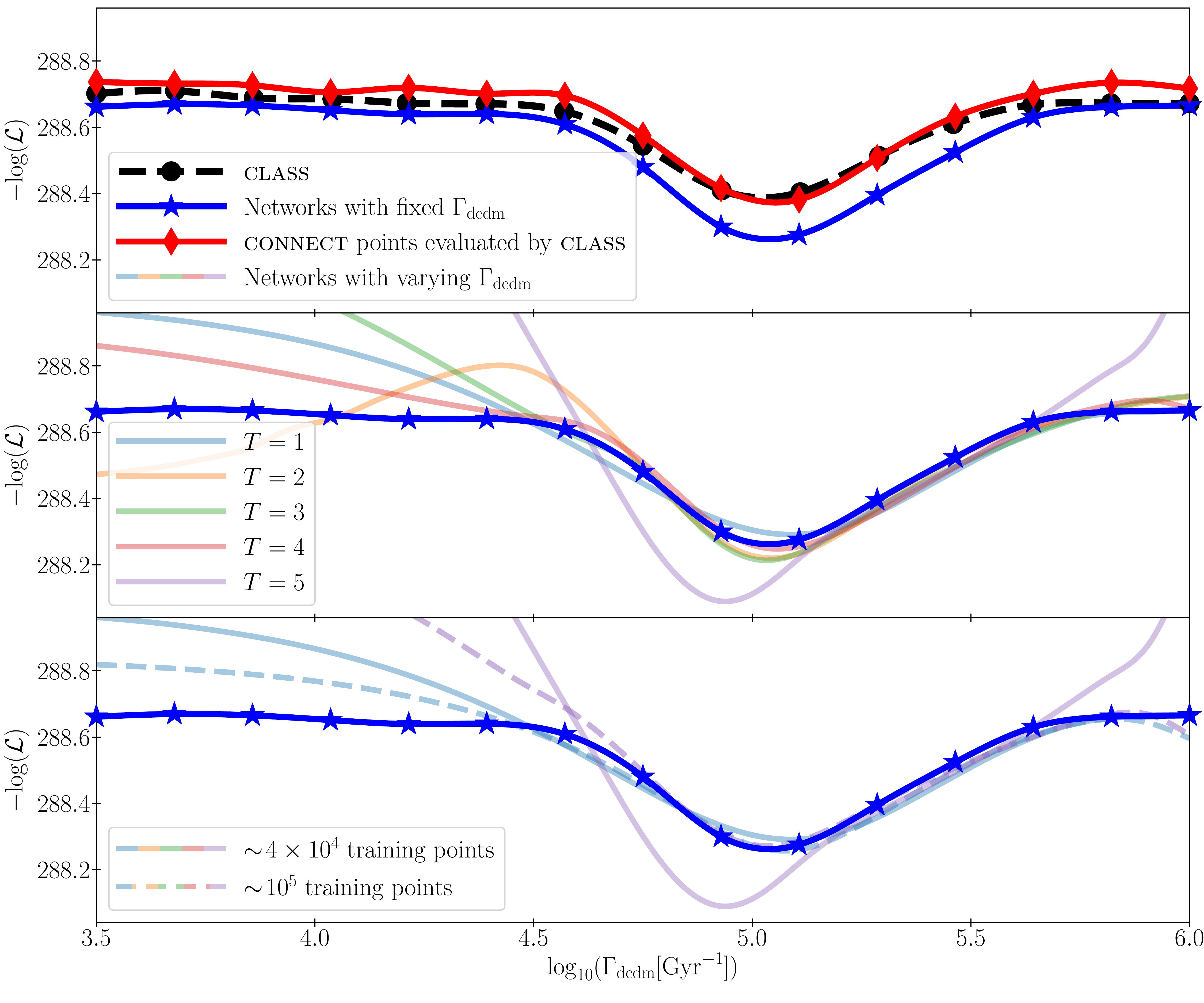}
	\caption{\textsl{The top panel shows profile likelihoods in the $\Gamma_{\rm dcdm}$ parameter with both \class{} (simulated annealing) shown with black circles and \connect{} (basin-hopping and BFGS) using several individual networks for each fixed value of $\Gamma_{\rm dcdm}$ shown with blue stars. The parameter vectors resulting from the \connect{} optimisations have all been evaluated by \class{} and plotted as a profile likelihood, shown with red diamonds. The middle panel shows profile likelihoods resulting from neural networks with $\Gamma_{\rm dcdm}$ as a varying parameter where training data has been gathered at different sampling temperatures (see text). The bottom panel shows the same kinds of profiles as the middle panel, but with different amounts of training data for the networks. Only the temperatures T=1 and T=5 are shown in order to keep the figure simpler.}}
	\label{fig:gamma_dcdm}
\end{figure}

Figure~\ref{fig:gamma_dcdm} shows the profile likelihood in the $\Gamma_{\rm dcdm}$ parameter using both \class{} and \connect{}. \connect{} has been tested both using single networks trained on the entire parameter space at different sampling temperatures and individual networks for each fixed value of $\Gamma_{\rm dcdm}$. The individual networks for each fixed value of $\Gamma_{\rm dcdm}$ have been trained with the regular iterative approach of \connect{}~\cite{Nygaard:2022wri}, and the networks spanning the entire parameter space have been trained with a slight modification. The first part of the training is identical to the regular iterative approach with the default sampling temperature of $T=5$, but when convergence is reached, the sampling and training do not halt. Instead, new data is gathered with a smaller temperature ($T=4$) and used to train a new network, which is then used to gather data for a new network with an even lower temperature, etc. The training data from consecutive iterations are not merged in this second part since they have different statistical properties due to differences in sampling temperature. This annealing results in multiple networks with different sampling temperatures according to a predefined list that would not be possible to obtain with just the regular iterative approach. This is because the likelihood is difficult to sample with small temperatures, so having only a small temperature from the beginning will not result in a useable network for e.g.\ $T=1$. By having the data gathered with a slightly larger temperature as the foundation for the network gathering the next data with a smaller temperature, it ensures that the network is always very accurate when gathering data. The reason for wanting to do this is that we cannot properly resolve this particular best-fit region with too large of a temperature given that it is a very narrow and slight band in the multidimensional likelihood surface. On the other hand, we cannot perform the entire sampling at a low temperature because the accuracy would suffer. With inspiration from the simulated annealing algorithm, this new approach takes all of this into account. 

Another problem apparent in figure~\ref{fig:gamma_dcdm} is that the networks with different temperatures only seem to agree just around the minimum and for higher values of $\Gamma_{\rm dcdm}$\footnote{The network with $T=5$ is worse than the others due to it having training data from all previous iterations as well, which usually is not a problem, but in this case the precision needs to be very high.}. This is, however, not surprising since we are sampling training data from the posterior, which is heavily influenced by volume effects towards high values of $\Gamma_{\rm dcdm}$ in this case. We therefore have much more training data on the right side of the figure, which means that all of the networks have better performance here. 

The figure also shows results from individual networks with fixed values of $\Gamma_{\rm dcdm}$, and we can clearly see that these resemble the results from \class{} much more. There is a small discrepancy in the depth of the well, but this is due to small precision errors. The difference corresponds to $\Delta\chi^2\approx0.2$ and this is definitely not significant given that the same behaviour and shape are produced. The networks with different temperatures also seem to agree with the networks with fixed values of $\Gamma_{\rm dcdm}$ for high values, so it is highly probable that it is a systematic error founded in the precision of neural networks. These networks have, however, all been trained to per mille precision in all of the $C_{\ell}$ spectra, but when investigating the reason for the differences in $\chi^2$ we found that the Planck lite likelihood is very sensitive to certain ranges in $\ell$ for the $TE$ spectrum. A per mille error in the $TE$ spectrum can lead to errors in $\chi^2$ up to roughly $0.3$ around the best-fit, and that seems to be in agreement with what we can see from the figure. This is only relevant for this very slight likelihood peak, and since the peak is more significant when using the full Planck 2018 likelihood~\cite{Holm:2022kkd}, the precision is most likely not an issue. Increasing the precision in the networks, e.g.\ by training for more epochs, could potentially render the effect unnoticeable in figure~\ref{fig:gamma_dcdm}, but this would be without merit due to the $\chi^2$ precision already being much greater than what is needed in any actual case of use. Regardless of the minute $\chi^2$ discrepancy, the \connect{} networks seem to find the same optimal parameter vectors that \class{} finds, given that evaluating the points with \class{} results in more or less the same profile as \class{} finds on its own (red diamonds in the top panel of the figure). The $TT$ spectrum is furthermore known to have much more constraining power than both $TE$ and $EE$, so we should expect the same shape in the profile likelihoods due to the sufficiently high precision in the $TT$ spectrum, and the small discrepancy from the precision in the $TE$ spectrum only shifts the $\chi^2$ values by a small amount and not the parameter vectors. Parameter inference is therefore not affected by this discrepancy. This is supported by the curve with the red diamonds in figure~\ref{fig:gamma_dcdm}, which shows the same points in the parameter space obtained by the \connect{} networks with fixed values of $\Gamma_{\rm dcdm}$ but evaluated by \class{}. We see that this curve is very close to the one actually obtained by \class{}, and this indicates that the optimisations of the neural networks find the correct optima.

It stands to reason that we should be able to obtain the same level of accuracy with a single network spanning the entire parameter space as we can with individual networks at fixed values of $\Gamma_{\rm dcdm}$, and indeed the results seem to converge if we generate larger amounts of training data, which is apparent from the bottom panel of figure~\ref{fig:gamma_dcdm}. When increasing the amount of data much further than shown here, it is necessary to train the networks for more epochs and perhaps increase the size of the networks, i.e.\ hidden layers and nodes in each layer. Given that we have 15 individual fixed-value networks each with $\sim$$5\times10^4$ training points, 6 hidden layers, and 1024 nodes in each hidden layer, we can definitely justify making a much larger network instead with close to $\sim$$10^6$ training data points since this would be equally expensive in terms of computational power. Increasing the number of trainable parameters in the network also increases the amount of information it can hold, which eventually will bring the emulation error down to a point where there will be no discrepancy between \class{} and \connect{}.

\subsection{Computational performance}
It is obvious that the computational cost of computing profile likelihoods with \connect{} is much lower than when using \class{} simply because the evaluation time of a \connect{} network is around 3 orders of magnitude faster than that of \class{}~\cite{Nygaard:2022wri}. Just using \connect{} instead of \class{} with simulated annealing would therefore be a huge speed-up. The utilisation of gradients, however, boosts the speed-up even further since fewer function evaluations are needed. In order to use gradients in \tf{}, a computational graph of the entire gradient computation needs to be constructed, which roughly takes around a minute, after which the evaluation of both the likelihood and the gradients takes around $\sim$$10^{-2}$ seconds. Given that \class{} is allowed to run on multiple threads, the evaluation time is around $\sim$$1$ second. Using the basin-hopping algorithm described in section~\ref{sec:implementation} with the BFGS optimiser as the local optimiser, each local optimisation requires $\sim$$10^2$ function calls, each ensemble contains $\sim$$10^1$ walkers, and the temperature is updated until convergence (usually around 2--4 times). This results in $\sim$$10^3$--$10^4$ evaluations, each taking $\sim$$10^{-2}$ seconds. This means that a single point will usually converge in less than a minute on a single CPU core. The simulated annealing algorithm requires on the order of $\sim$$10^4$--$10^5$ evaluations for each point in the profiles presented in this paper, and with an evaluation time dominated by \class{}, a single point will converge in roughly a few days if run sequentially. There is, however, some degree of parallelisability, given that multiple chains can be utilised and \class{} can be parallelised to around 8--10 cores. The different points of a profile can, of course, be computed completely separately, regardless of which algorithm is used.

There is also the matter of gathering training data for a neural network if one is starting from scratch, and this process is quite time-consuming compared to the use cases of a trained network~\cite{Nygaard:2022wri}. For most likelihoods, the network requires around $\sim$$50,000$ points of training data, which means that \class{} needs to be evaluated $\sim$$50,000$ times. This is, however, very parallelisable, and it is much faster than doing an actual MCMC run with \class{} -- especially for beyond $\Lambda$CDM where the number of \class{} evaluations can be as high as $10^5$--$10^6$. If one only seeks to optimise a single point of a ($\Lambda$CDM) model, it might be better to use \class{} and simulated annealing, but for an entire one-dimensional profile, it is very beneficial to use \connect{} instead, and for two-dimensional profiles requiring several hundred points, it might be necessary to use \connect{}. If one seeks to perform a full frequentist analysis with triangle plots of one- and two-dimensional profiles, the task is virtually impossible at this point without using \connect{}.

\section{Conclusion and outlook}\label{sec:conclusion}

Using the ensemble basin-hopping algorithm for global optimisation combined with the BFGS optimiser for local optimisation has been demonstrated to yield robust, fast, and accurate results when calculating profile likelihoods from CMB data.
By making use of gradients in the local optimiser, the number of function evaluations can be greatly decreased when compared to gradient-free simulated annealing. This, together with the much faster evaluation time of \connect{} compared to \class{}, results in speed-ups of several orders of magnitude when constructing profile likelihoods. In addition to being fast, the method is also very robust and accurate, and given the smoothness of the neural networks, the global optimisations often converge with only a few local optimisations. 

When a neural network has been trained, the profile likelihoods in any parameter or set of parameters are computationally very inexpensive, and entire triangle plots of profile likelihoods, typically consisting of more than 5000 individual points in parameter space for which optimisation must be performed,
are easily computed. Each such point in the parameter space is independent of all other points, making the optimisations embarrassingly parallelisable. The optimisation of each point takes around a minute on a single modern CPU core. A full triangle plot could therefore be computed on a normal quad-core laptop in less than a day.

With fast and easy access to profile likelihoods in cosmology, it is easy to investigate cosmological models with both Bayesian and frequentist statistics. Neither of these statistical frameworks can claim superiority compared to the other, but the two different approaches answer different questions and complement each other. It is therefore very useful to have both a posterior and a profile likelihood in order to get the full picture and draw reasonable conclusions about the given model. The biggest reason for this not being done frequently in analyses of cosmological models is that the computational costs of profile likelihoods are much greater than those of posteriors. Having a fast emulation tool for quick computations of both posteriors and profile likelihoods makes it much easier and more appealing to (re)introduce the frequentist approach in cosmology.

While the optimisation is typically extremely precise, the precision is limited by the emulation precision of the underlying network. Therefore, it is quite important to use well-trained and precise networks to derive profiles consistent with those obtained using \class{} and simulated annealing.
In this work, the same neural network has been used for all profile likelihoods and MCMC related to the $\Lambda$CDM model and the massive neutrino model due to its simple shape and only modest volume effects. The agreement with profile likelihoods from \class{} is excellent for these two models and very precise out to several standard deviations in parameter space. For the DCDM model, the agreement with \class{} is also quite good when using either individual networks for each point in the decay rate, $\Gamma_{\rm dcdm}$, or an annealed network with a large amount of training data, even though a small offset is visible between the profile likelihoods. This is, however, on a scale of $\Delta\chi^2\approx0.2$ which is much too small to be significant, and parameter inference would not be affected by this small discrepancy. 

The training data for the \connect{} neural networks are sampled iteratively from the posterior, and with large volume effects, this can bias the accuracy of the network away from regions of maximal likelihood: Regions with a large volume are sampled much more than regions with a small volume, even though the small volumes might have better likelihood values. This creates a bias in the networks towards larger volumes due to these being much more represented in the training data. It is possible to overcome this problem using a larger network architecture, more training data, and more training epochs. The training data can also be weighted in a way that reduces the impact of data points with larger values of $\Gamma_{\rm dcdm}$ on the total loss during training. However, it might be beneficial to pursue new ways of sampling training data more suited for profile likelihoods in future works.

Another thing to consider is the complexity of the likelihood function at this point. We evaluate a network to get $C_\ell$ spectra, perform a cubic interpolation, and use a \tf{} version of the Planck lite likelihood in order to go from cosmological parameters to a likelihood value, and this makes the computational graph of the gradients quite large in terms of memory. A way to speed up the computations even further and increase usability is to directly emulate the likelihood value of any given likelihood code. We then lose the dependency on having likelihood codes written in \tf{} syntax or emulated separately. Most of the problems highlighted in this paper will most likely cease to exist with this approach of directly emulating likelihoods.
An additional advantage is that any likelihood emulated in this way (e.g.\ BAO, SNI-a, etc.) will automatically lend itself to auto-differentiation, allowing the efficient combination of basin-hopping and BFGS local optimisation to be used. \\

\noindent {\bf Reproducibility.}
We have used the publicly available \connect{} framework available at \url{https://github.com/AarhusCosmology/connect_public} to create training data and train neural networks. The framework has been extended with the basin-hopping optimiser and a module for computing profile likelihoods. Explanatory parameter files have been included in the repository in order to easily use the framework and reproduce results from this paper.

\section*{Acknowledgements}
We acknowledge computing resources from the Centre for Scientific Computing Aarhus (CSCAA). A.N., E.B.H., and T.T. were supported by a research grant (29337) from VILLUM FONDEN.


\appendix
\section{Choosing points for two-dimensional profile likelihoods}\label{app:2d}
When computing two-dimensional profile likelihoods, we often need many more points than what is suitable for one-dimensional profile likelihoods, especially if the points are not selected in a clever way. A square grid with the same ranges as the one-dimensional profile likelihoods is certainly possible, but not the most feasible in terms of computational resources. Ideally, we would like to choose a collection of points from the best-fit region stretching to at least the $3\sigma$ contours (99.73\% confidence level) in order to have enough points to accurately compute the 68.27\% and 95.45\% confidence regions. A reasonable choice of points is choosing the bin centres of the histogram of an MCMC run (using e.g.\ \montepython{}~\cite{Audren:2012wb,Brinckmann:2018cvx}) with the same \connect{} model where the bin counts are different from zero. This means that each point will correspond to a small region where the MCMC sampler has accepted at least one point. This is what is used to compute the two-dimensional credible regions of the posterior, and the idea is that the same points should well represent at least part of the confidence regions of the profile likelihood. This will almost always be the case unless very significant volume effects are at play. 

\begin{figure}[t]
	\centering
	\includegraphics[width=\textwidth]{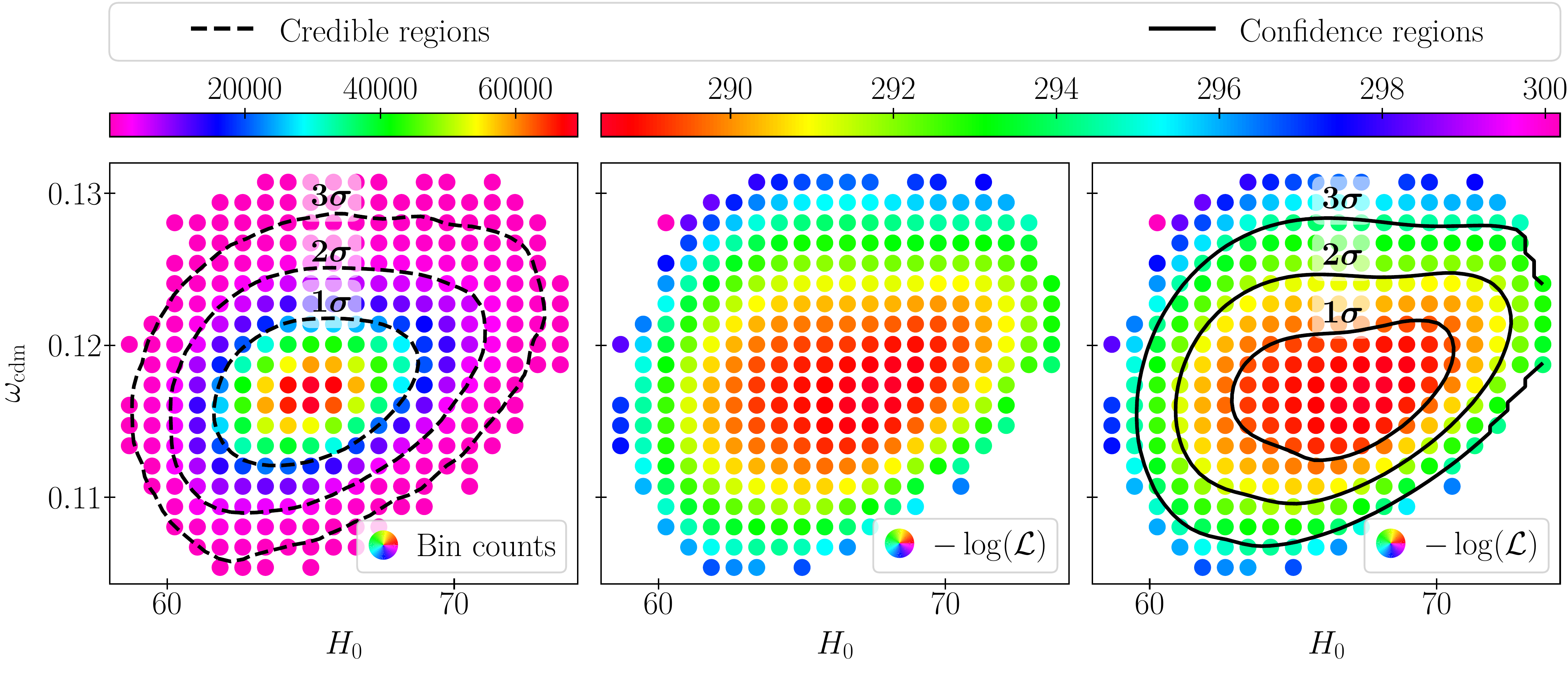}
	\caption{\textsl{Bin centres from the histogram in the ($\omega_{\rm cdm}$--$H_0$)-plane from \montepython{}. The left panel is coloured according to bin counts of the histogram, and the dashed $1\sigma$, $2\sigma$, and $3\sigma$ contours are the 68.27\%, 95.45\%, and 99.73\% credible regions, respectively. The middle and right panels are coloured according to the likelihood values, and the solid $1\sigma$, $2\sigma$, and $3\sigma$ contours in the right panel are the 68.27\%, 95.45\%, and 99.73\% confidence regions, respectively. The points from the histogram are suitable for the 68.27\% and 95.45\% confidence regions, but additional points are required for the 99.73\% confidence region.}}
	\label{fig:hist}
\end{figure}

Figure~\ref{fig:hist} shows the bin centres of one such histogram, and the different panels are coloured according to either the bin counts or the likelihood value. We can see that the points encapsulate the entire 99.73\% credible region but not the entire 99.73\% confidence region. The 68.27\% and 95.45\% confidence regions are, however, well represented. More elaborate methods for obtaining points in a clever way could be employed, but it is worth mentioning that this approach guarantees that the optimisation is embarrassingly parallelisable. 

\section{Recomputing and adding points}\label{app:addfix}
\begin{figure}[t]
	\centering
	\includegraphics[width=\textwidth]{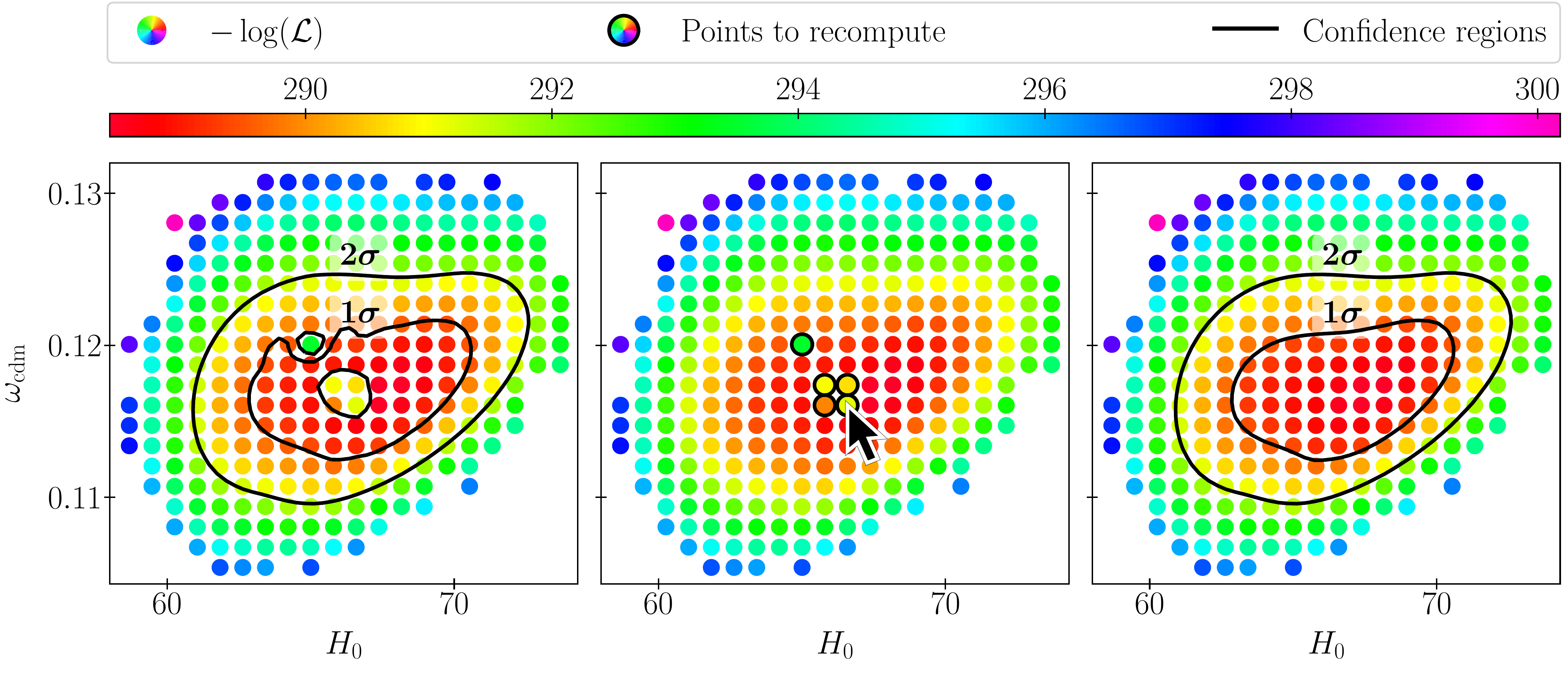}
	\caption{\textsl{The panels from left to right show the process of recomputing specific points in the two-dimensional ($\omega_{\rm cdm}$--$H_0$)-profile likelihood if the optimiser converges on a local optimum. The points to recompute can be selected interactively and are then optimised again with a better result.}}
	\label{fig:fix}
\end{figure}
\begin{figure}[t]
	\centering
	\includegraphics[width=\textwidth]{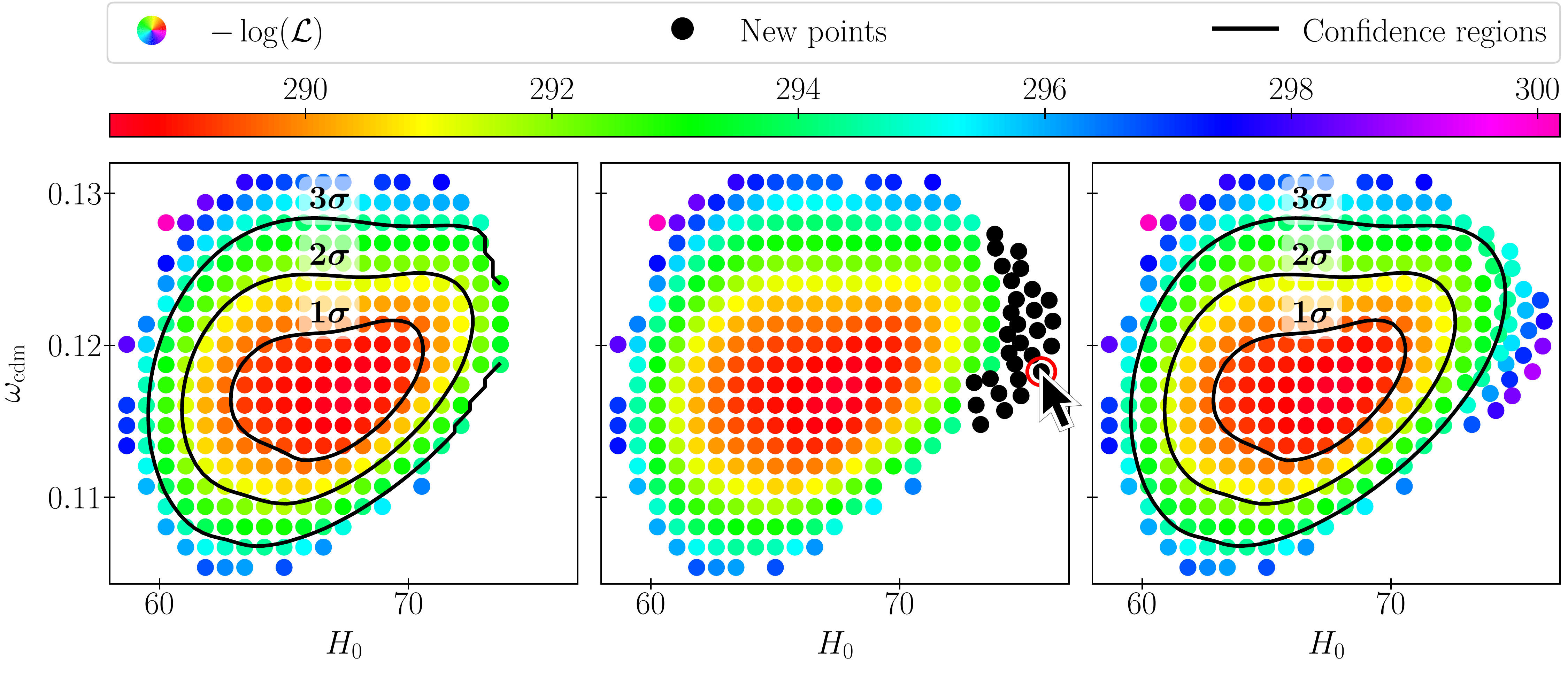}
	\caption{\textsl{The panels from left to right show the process of adding extra points to the two-dimensional ($\omega_{\rm cdm}$--$H_0$)-profile likelihood if the set of points does not encapsulate the entire contour of a confidence level. New points can be chosen interactively and are then optimised to complete the $3\sigma$ contour.}}
	\label{fig:add}
\end{figure}

Since the optimisation routine, as described in section~\ref{sec:implementation}, has some level of stochasticity, the optimisation might fail by converging on a local optimum a small fraction of the time. By tweaking precision settings and hyper-parameters, the rate of failed optimisations can be greatly decreased, but there will always be some probability of not succeeding. With the default settings, a complete triangle plot of profile likelihoods containing $\sim$$10^3$ points to compute will result in only $\sim$$10^1$ failed points. This is difficult to detect automatically, but it is very easily seen when plotting the profile likelihoods. A routine to interactively choose failed points after plotting them has been implemented, and figure~\ref{fig:fix} shows the process of choosing points to recompute. These are then gathered in a file and recomputed. Given the low probability of getting stuck in a local optimum, the recomputation is almost always successful. In rare cases, a few points need to be recomputed twice, and if a specific point turns out to be particularly difficult to optimise, then the precision settings might need adjustment for that single optimisation. This is still much more feasible than running with very precise settings for all points in the profile likelihoods.

If the computed points, chosen according to the process described in appendix~\ref{app:2d}, do not encapsulate the contour of a specific confidence level, additional points have to be selected and optimised. This is also difficult to choose automatically, but it is very easy to pick new points by looking at the contours and previously optimised points. A routine for interactively choosing new points has also been implemented, and using this, one can easily choose points based on the location of all current points and the contours based on those. Figure~\ref{fig:add} shows the process of choosing new points since the 99.73\% confidence region is not entirely encapsulated by the previously chosen points. The new points are gathered in a file and optimised. After the inclusion of the new optimised points, the contour line looks as it should and is fully represented by the total set of points.

\bibliographystyle{utcaps}
\bibliography{connect2023}

\end{document}